\documentclass[aps,prl,preprint]{revtex4-2}
\usepackage{graphicx}  % needed for figures
\usepackage{dcolumn}   % needed for some tables
\usepackage{bm}        % for math
\usepackage{amssymb}   % for math
 \usepackage{float}
\usepackage{amsfonts}
\usepackage{gensymb}
\usepackage{amsmath}
\usepackage{upgreek}
\usepackage{svg}
 \usepackage{hyperref}

\hypersetup{
   colorlinks=true,
    linkcolor=blue,
    filecolor=magenta,      
    urlcolor=blue,
}
 
\usepackage{blindtext,tikz}
\usetikzlibrary{calc}
\begin{document}
\title{Analogous Black Holes in Type-III Dirac Semimetal Ni$_3$In$_2$X$_2$ (X = S, Se)}

\author{Christopher Sims}
\affiliation{\textit{School of Electrical and Computer Engineering, Purdue University, West Lafayette, IN 47907, USA}}
\date{\today}

\begin{abstract}
Black holes are objects that have a large mass and curve space time, characterized by their event horizon and singularity. Recently, an interesting concept of analogous black holes has emerged in the field of condensed matter physics. In this work, the possibility of realizing analogous black holes in topological material is Ni$_3$In$_2$X$_2$ (X = S, Se) discussed. This work shows that the type-III Dirac cones of the material can lead to the emergence of an event horizon and the formation of a black hole-like region near the Dirac point. In addition, possible experimental signatures of such a system are discussed and the potential implications of an analogous black hole for the study of black hole physics in condensed matter systems.
\end{abstract}
\date{\today}
\maketitle
\section{Introduction}
Black holes are extreme objects in the universe, characterized by their large gravitational pull, event horizon, and singularity \cite{Hawking1974, Bekenstein1973}. Black holes have been the subject of intense study in physics and astrophysics due to their unique properties and their potential implications in understanding the nature of gravity and the universe as a whole.

Recently, the concept of analogous black holes has emerged in the field of condensed matter physics \cite{Barcelo2005, Unruh1981}. Analogous black holes are not actual black holes but systems that mimic some of the properties of black holes, such as an event horizon and a Hawking-like radiation \cite{Knott2018, Cardoso2016}. The emergence of analogous black holes in condensed matter systems has been a subject of intense research in recent years, with several proposals and experimental realizations. The study of photon interactions has been seen to have analogous effect to a black hole horizon \cite{Steinhauer2016, Belgiorno2010}. More recently, topological systems have been proposed to have ideal states for the study of analogous gravitational interactions in black holes \cite{Nikitin2019,Krishna2020,Jacobson1998,Jacobson2002,Jacobson1998a,Volovik1999,Huhtala2002,Nissinen2020,Nissinen2017}, or for analogous wormholes \cite{Dai2019,Simonetti2020,Sims2021}

Topological materials are a class of materials characterized by their nontrivial band topology, which gives rise to a range of exotic electronic properties \cite{Hasan2010, Qi2011}, such as Weyl\cite{Xu2015,Lv2015,Huang2015,Weng2015,Wan2011} and  Dirac\cite{Borisenko,Wang2013,Neupane2014,Neupane2016} semi-metals. It is shown that topological properties of the material can lead to the emergence of an event horizon and the formation of a black hole-like region in the system within the SYK model \cite{Rosenhaus2019}. Experimental signatures of such a system and explore the potential implications in the field of condensed matter physics.

Recently, a new type of Dirac semimetal phase, called type-III, has been predicted to exist in some materials \cite{Fragkos2021,Milicevic2019}. In type-III Dirac semimetals, the Dirac cones are tilted, leading to an anisotropic dispersion relation. This results in an unusual electronic structure that exhibits a number of exotic properties, such as chiral anomaly and topological Lifshitz transitions \cite{Mizoguchi2022,Volovik2017}.

In this paper, Topological materials Ni$_3$In$_2$S$_2$ and Ni$_3$In$_2$Se$_2$ are studied, which have recently been found to exhibit an endless nodal-line phase near the Fermi level\cite{Zhang2022}. Using first-principles calculations, The electronic structure of Ni$_3$In$_2$S$_2$ and Ni$_3$In$_2$S$_2$ are investigated and presence of a type-III Dirac semimetal phase is found in the band structure.

\section{Materials and Methods}
The band structure calculations were carried out using the density functional theory (DFT) program Quantum Espresso (QE) \cite{Giannozzi2009} with GPU acceleration on CUDA version 11.7. Generalized gradient approximation (GGA) \cite{pbegga} was set as the exchange correlation functional. Projector augmented wave (PAW) pseudo-potentials were generated utilizing PSlibrary \cite{Corso2014}. The relaxed crystal structure was obtained from materials project \cite{Jain2013, Sims2020} for Ni$_3$In$_2$S$_2$ and Ni$_3$In$_2$S$_2$. The relaxed crystal parameters are used to calculate the band structure for Ni$_3$In$_2$S$_2$[Table~\cite{supp}], and Ni$_3$In$_2$Se$_2$ [Table~\cite{supp}]. The energy cutoff was set to 100 Ry (1360 eV) and the charge density cutoff was set to 400 Ry (5442 eV) for the plane wave basis, with a k-mesh of 25 $\times$ 25 $\times$ 25. High symmetry point K-path was generated with SSSP-SEEK path generator \cite{Hinuma2017,Togo2018}. The bulk band structure [Fig.~\ref{CISband}] was calculated from the 'SCF' calculation by utilizing the 'BANDS' flag in Quantum Espresso. As oppose to utilizing plotband.x included in the QE package, a custom python code is used to plot the band structure with the matplotlib package. Topological number analysis was conducted with the Wilson loop algorithm contained within the WannierTools package. 

Single crystals of Ni$_3$In$_2$S$_2$ and Ni$_3$In$_2$S$_2$ (SG: R$\overline{3}$M [166]) are grown via the Indium flux method \cite{Fisk1989,Canfield1992}. Stoichiometric quantities of Ni (99.9\%, Alfa Aesar)  and Se ($\sim$200 mesh, 99.9\%, Alfa Aesar)/ S ($\sim$325, 99.5\%, Alfa Aesar) were mixed and ground together with a mortar and pestle. Indium (99.99\% RotoMetals) was added in excess (50\%) in order to allow for a flux growth. All precursor materials were sealed in a quartz tube under vacuum and placed inside a high temperature furnace. The sample was heated up to 1000 $^{\circ}$C over 1440 min, kept at 1000 $^{\circ}$C for 1440 min, cooled down to 950 $^{\circ}$C over 180 min, kept at 950 $^{\circ}$C for 2880 min, then slowly cooled down to 180 $^{\circ}$C where the sample was taken out of the furnace then centrifuged. These growth parameters were also used to grow Co$_3$In$_2$X$_2$ (X=S,Se) single crystals. The grown crystals characterized via LEED (OCI LEED 600) and powder X-ray diffraction (XRD) (Bruker D8 DISCOVER, Cobalt Source) to confirm their crystal structure. The XRD results and calculations have been normalized, this results in some features in the calculated XRD being more prominent. [See Supplementary for XRD, LEED~\cite{supp}]

%\begin{figure*}[ht]
% \centering
% \includegraphics[width=1.0\textwidth]{RMGARC.png}
%  	\caption{\textbf{Fermi Surface of RMn$_6$Ge$_6$:} (A) NdMn$_6$Ge$_6$ (NMG) (B) SmMn$_6$Ge$_6$ (SMG) (C) TbMn$_6$Ge$_6$ (TMG) (D) DyMn$_6$Ge$_6$ (DMG) (E) HoMn$_6$Ge$_6$ (HMG) (F) ErMn$_6$Ge$_6$ (EMG) (G) YbMn$_6$Ge$_6$ (YMG) (H) LuMn$_6$Ge$_6$ (LMG)}
%\label{ARC}
%  \end{figure*}

%\begin{figure*}[ht]
% \centering
% \includegraphics[width=1\textwidth]{RMGSS}
%  	\caption{\textbf{Surface states along the $\mathrm{\overline{\Gamma}}$-$\mathrm{\overline{K}}$-$\mathrm{\overline{M}}$-$\mathrm{\overline{K}}$-$\mathrm{\overline{\Gamma}}$ line:} (A) NdMn$_6$Ge$_6$ (NMG) (B) SmMn$_6$Ge$_6$ (SMG) (C) TbMn$_6$Ge$_6$ (TMG) (D) DyMn$_6$Ge$_6$ (DMG) (E) HoMn$_6$Ge$_6$ (HMG) (F) ErMn$_6$Ge$_6$ (EMG) (G) YbMn$_6$Ge$_6$ (YMG) (H) LuMn$_6$Ge$_6$ (LMG)}
%\label{SS}
% \end{figure*}

\section{Results and Discussion}
\subsection{SYK Model}
The Sachdev-Ye-Kitaev (SYK) model is a quantum many-body system that is exactly solvable. the SYK is able to accurately model 2D gravity, which is ideal for an accurately solvable model for black holes.
where $\psi_i$ are Majorana fermions, $J_{ijkl}$ are random couplings with a Gaussian distribution. O is used to capture higher order terms \cite{Sachdev1993,Polchinski2016,Maldacena2016}.

\begin{equation}
H = \sum_{i<j<k<l} J_{ijkl} \psi_i \psi_j \psi_k \psi_l - \frac{\mu}{4!} \sum_{i<j<k<l} \psi_i \psi_j \psi_k \psi_l + O
\end{equation}

In the SYK model, the $\mu$ parameter is called the ``mass" term. In the context of black holes, the parameter $\mu$ is related to the temperature and entropy of the black hole, and plays an important role in the calculation of thermodynamic properties using the SYK model. 

Contrarily, for normal interacting Dirac Fermions in 2D, the SYK model takes the form\cite{Bernard2002}:
\begin{equation}
H = \sum_{i<j<k<l}J_{ijkl} \psi_{i}^{\dagger} \psi_{j} \psi_{k}^{\dagger} \psi_{l}- \frac{\mu}{4!}\sum_{i<j<k<l} \psi_{i}^{\dagger} \psi_{j} \psi_{k}^{\dagger} \psi_{l} + O
\end{equation}

where $\psi_i$ are Dirac Fermions, $J_{i_1 i_2 \dots i_q}$ are random couplings with a Gaussian distribution $q$ is the order of the interaction. For the critically tilted regime, a term for the dispersion of the Dirac Fermions is added \cite{Seo2022,Landsteiner2016,Gao2023}.

\begin{equation}
H = \sum_{i<j<k<l}J_{ijkl} \psi_{i}^{\dagger} \psi_{j} \psi_{k}^{\dagger} \psi_{l} + \sum_{i<j} t_{ij} \psi_{i}^{\dagger} \psi_{j}  - \frac{\mu}{4!}\sum_{i<j<k<l} \psi_{i}^{\dagger} \psi_{j} \psi_{k}^{\dagger} \psi_{l}+ O
\end{equation}

where $\psi_i$ are Dirac fermions, $J_{ijkl}$ are random couplings with a Gaussian distribution, $t_{ij}$ is a hopping parameter that describes the band structure of the Dirac cone and $\mu$ is the chemical potential. Since all of these models are exactly solvable, it is possible to measure the dispersion and interaction of Fermions (or Majorana Fermions) in these models. It is important to note that it is only in the critically tilted regime for Dirac Fermions where $\psi_{i}^{\dagger} = - \psi_{j}$, thusly Dirac Fermions behave like Majorana Fermions \cite{Gao2023}. Therefore, pseudo Black hole interactions can be probed in condensed matter systems.

\begin{figure*}[!ht]
 \centering
 \includegraphics[width=1.0\textwidth]{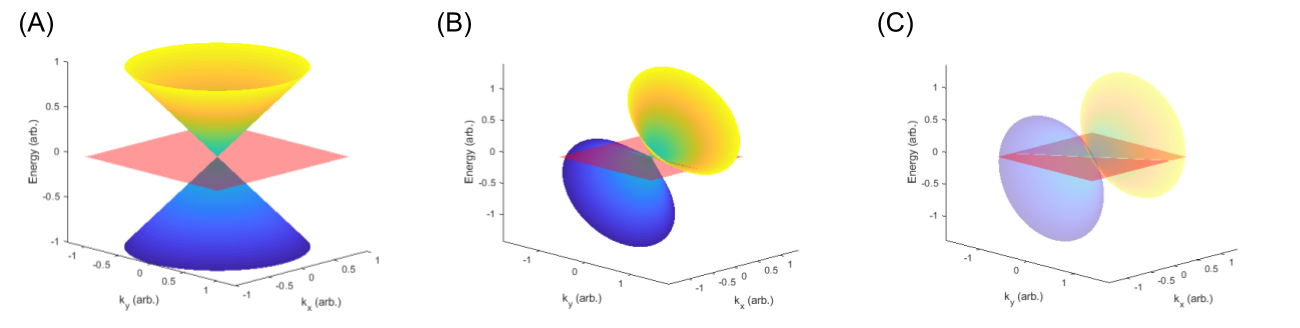}
  	\caption{\textbf{Tilted Dirac Cone}: An illustration of the tilted Dirac cone in its (A) type-I normal phase (B) critically tilted Type-III phase, and its (C) over-tilted type-II Phase }
\label{cone}
 \end{figure*}
The Dirac cone can have three main phases, its type-I normal phase [Fig \ref{cone}(A)], its critically tilted type-III phase  [Fig \ref{cone}(B)], and its over tilted phase [Fig \ref{cone}(C)]. By changing the interaction terms in the SYK model, it is possible to model all of these Dirac cone phases within the toy model with 4 interaction terms and only 2D hoping parameters. Within this work, only the type-III phase is of interest.

\subsection{Bulk band structure of Ni$_3$In$_2$X$_2$(X = S, Se)}
\begin{figure*}[!ht]
 \centering
 \includegraphics[width=1.0\textwidth]{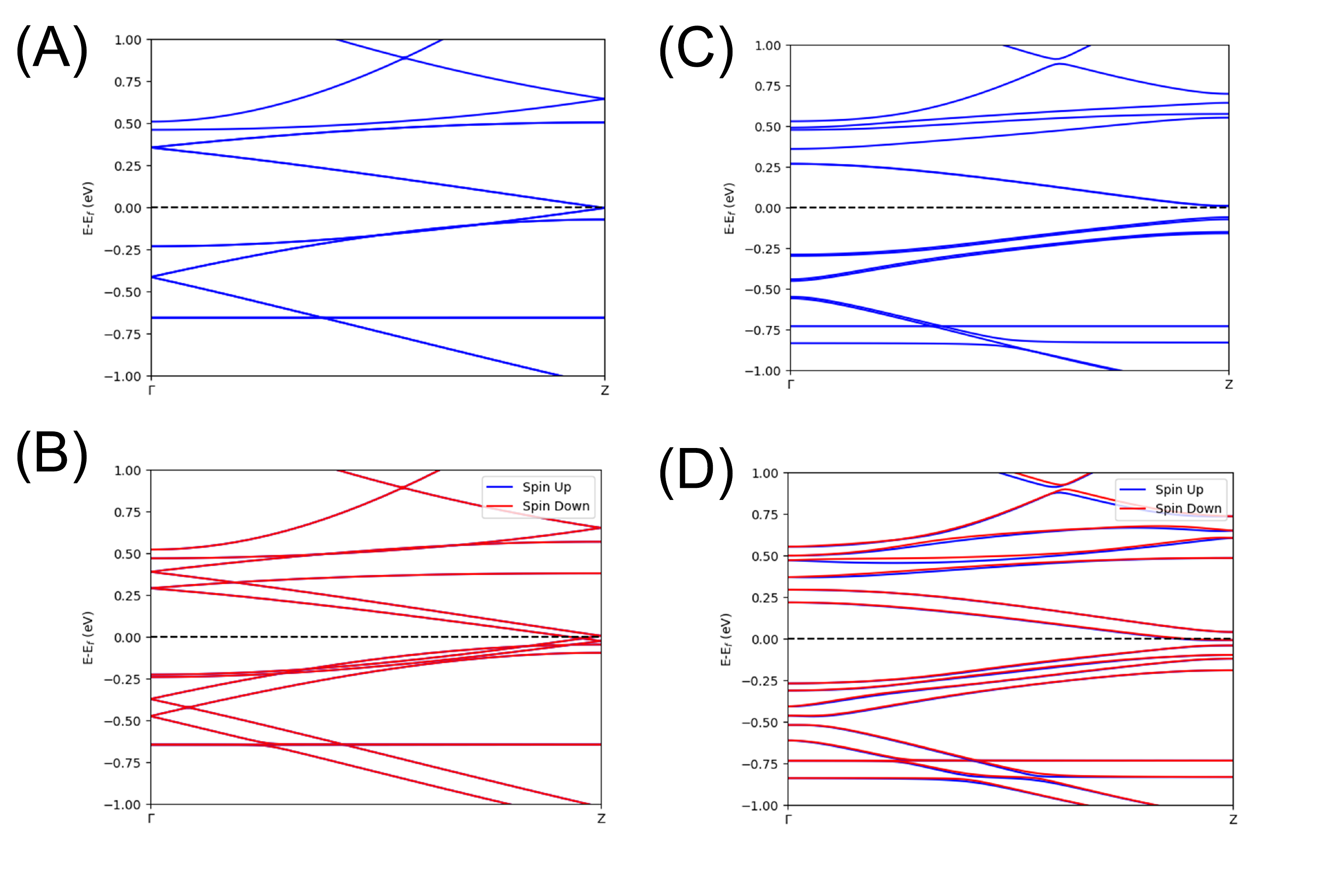}
  	\caption{\textbf{Bulk Bands structure}: The bulk band structure of Ni$_3$In$_2$S$_2$ along the $\Gamma$-Z high symmetry line (A) without SOC (B) with SOC. For Ni$_3$In$_2$Se$_2$ (C) without SOC (D) with SOC}
\label{bulk}
 \end{figure*}

Single crystals of Ni$_3$In$_2$X$_2$(X = S, Se) and analyzed with powder XRD and are found to have good agreement with their predicted crystal structure [see supplementary XRD ~\cite{supp}]. The calculated crystal parameters are a = b = 5.432 $\mathrm{\AA}$, c = 13.606 $\mathrm{\AA}$ for Ni$_3$In$_2$Se$_2$ and a = b = 5.377 $\mathrm{\AA}$, c = 13.482 $\mathrm{\AA}$ for Ni$_3$In$_2$S$_2$ [see supplementary for lattice site positions~\cite{supp}]. The single crystal nature of Ni$_3$In$_2$X$_2$(X = S, Se) is also analysis with LEED measurements and are found to have perfect hexagonal shape associated with the R$\overline{3}$M space group.

Bulk band calculations show that there is a flat band in the $\Gamma$-Z direction about -700 meV below the Fermi level for Ni$_3$In$_2$S$_2$  [Fig \ref{bulk}(A)], and two flat bands -750 meV and -800 meV Below the Fermi level for Ni$_3$In$_2$Se$_2$  [Fig \ref{bulk}(C)]. Detailed analysis shows that these bands are perfectly flat to within 4 digits $\pm 0.00005$ eV , $\pm 0.05$ meV for the entire high symmetry line (where 0.0001 is the limit of the accuracy). However, the region of interest are the type-III Dirac cones that form in the gap when spin orbit coupling is applied. Ni$_3$In$_2$S$_2$  opens two type-III Dirac cones with a gap $\sim 30$ meV which are iso-energy and are connected by the flat bulk band [Fig \ref{bulk}(B)]. Ni$_3$In$_2$S$_2$  opens two sets of type-III Dirac cones with a gap of $\sim 30$ meV for the upper two cones, and a gap of $\sim 100$ meV for the lower two Dirac cones [Fig \ref{bulk}(D)]. It is important to note that there is also a momentum ($K$) gap for the Dirac cones in the type-III case. In order to confirm the topological nature of the Dirac cones, $\mathbb{Z}_2$ analysis is conducted and it is found that there is nontrivial topology near the predicted spots for the Dirac cones. The topological number analysis shows that there are prominent areas which have a strong topological charge and only occur and certain points, leading to peaks in the loop analysis when typically smooth curves are expected, this is due to the type-III Dirac point having bands which exist on the same energy level and cross each other simultaneously causing divide by zero states, which is unique only to the type-III phase.

\subsection{Experimental probes}
Magneto-transport probes of type-III Dirac semi-metals has already been discussed in detail \cite{Mizoguchi2022}, therefore, another method of measuring an analogous black hole is provided.

Super-radiant (SR) scattering for black holes can be discribed by the Teukolsky equation\cite{Bini2002}. The Teukolsky equation is not an exactly solvable model and must be solved via numerical methods. However, proposals have shown that SR scattering is an ideal test for general relativity near a black hole (where it breaks down) \cite{Yagi2016}. Dirac Fermions have been theoretically shown to also be an ideal test of SR scattering near a black hole in Kerr spacetime\cite{Dolan2015} where Kerr spacetime provides an ideal test bed for testing particle physics on a black hole \cite{Brito2015,Herdeiro2015}. An infrared (IR) laser is sent at an incidence angle of are 45$^{\circ}$ to single crystals of p-doped Ni$_3$In$_2$As$_{x}$Se$_{2-x}$ (with respect to the $\mathrm{\hat{C}}$ - axis). In order to probe only the ``black-hole'' side of the type-III Dirac cone, the laser should be circularly polarized by sending a laser that is linearly polarized through a quarter wave plate and an angle of 45$^{\circ}$. In order to detect the reflected light, and analyzer (polarizer) with a photodetector can be setup on the reflected side. By tuning the analyzer side, it is possible to measure polarized response of the photoelectrons from the `left' and `right' sides of the Dirac cone. SR scattering should show an incidence angle dependence and a polarization dependence to experimentally demonstrate the existence of an analogous black-hole in the type-III Dirac semi-metal Ni$_3$In$_2$Se$_2$

\section{Conclusion}
In conclusion, it is found that the SYK model, with modifications, can capture the interaction of both a black hole with Majorana Fermions and Dirac Fermions with similar equations. Upon studying the critically tilted Dirac cone, it can be seen that the Dirac Fermions behave in a similar manner to Majorana Fermions in the Black hole model. By utilizing this correlation it is possible to study black hole physics in condensed matter systems, opening up the possibility of studying quantum information paradoxes in a table top setting. Single crystals of Ni$_3$In$_2$S$_2$ and Ni$_3$In$_2$Se$_2$ have been shown theoretically to have perfectly flat and ideal type-III Dirac cones. Finally, an experiment is proposed to both confirm analogous black holes in type-III Dirac semimetals.

 \section{Acknowledgments}
 \vspace{-4mm}
C.S. Acknowledges the generous support from the GEM Fellowship and the Purdue Engineering ASIRE Fellowship.
 
 Correspondence and requests for materials should be addressed to C.S.
 (Email: Sims58@Purdue.edu)

 \section{Supplementary}
  \vspace{-4mm}
See supplementary for further details on crystal diffraction, LEED analysis and band structure calculations of Ni$_3$In$_2$X$_2$ (X = S, Se).

%\bibliography{NIS_REF}
%aipnum4-2.bst 2019-01-14 (MD) hand-edited version of apsrev4-1.bst
%Control: key (0)
%Control: author (8) initials jnrlst
%Control: editor formatted (1) identically to author
%Control: production of article title (-1) disabled
%Control: page (0) single
%Control: year (1) truncated
%Control: production of eprint (0) enabled
%

\end{document}